\documentclass[a4paper]{jpp}
\usepackage{graphicx}
\usepackage{epstopdf, epsfig}

\usepackage[normalem]{ulem}
\usepackage{amsfonts,amsmath,amssymb}
\usepackage{environ}
\usepackage{graphicx}
\usepackage[breaklinks=true,colorlinks=true,allcolors=blue]{hyperref}
\usepackage{breakurl}
\usepackage[leftcaption]{sidecap}
\usepackage{psfrag}
\newcommand{\mypsfrag}[2]{\psfrag{#1}{\scriptsize{#2}}}
\usepackage{tabularx}
\usepackage{ragged2e}
\newcolumntype{Y}{>{\RaggedRight\arraybackslash}X}
\usepackage{booktabs}
\usepackage{xspace}
\usepackage[UKenglish]{babel}
\usepackage{placeins}

\usepackage[]{natbib}
\bibpunct[; ]{(}{)}{,}{a}{}{;}
\usepackage{doi}
\usepackage{url}

\renewcommand{\vec}[1]{\mbox{\boldmath $ #1$}}
\newcommand{\bm}[1]{\mbox{\boldmath $ #1$}}
\renewcommand{\v}{\vec}

\newcommand{\B}{\vec B}
\renewcommand{\u}{\vec u}

\newcommand{\ru}{\hat{\vec r}}
\renewcommand{\r}{\vec r}
\renewcommand{\k}{\hat{\vec k}}

\newcommand{\rhobar}{\bar{\rho}}
\newcommand{\Tbar}{\bar{T}}
\newcommand{\Pbar}{\bar{P}}

\newcommand{\Nrho}{N_\rho}
\newcommand{\pol}{v}
\newcommand{\tor}{w}
\newcommand{\polB}{h}
\newcommand{\torB}{g}

\newcommand{\R}{\mathrm{R}}
\newcommand{\Ek}{\mathrm{Ek}}
\renewcommand{\Pr}{\mathrm{Pr}}
\newcommand{\Pm}{\mathrm{Pm}}

\newcommand{\Z}{{\mathrm{Z}}}

\usepackage{color}
\newcommand{\red}[1]{#1} 
\newcommand{\redmath}[1]{#1} 

\allowdisplaybreaks

\shorttitle{Flows and dynamos in radiation zones}
\shortauthor{R. D. Simitev and F. H. Busse}

\title{Flows and dynamos in a model of stellar radiative zones}

\author{Radostin
  D. Simitev\aff{1}\corresp{\email{\href{mailto:Radostin.Simitev@glasgow.ac.uk}{Radostin.Simitev@glasgow.ac.uk}}}\aunote{ORCID
    ID Radostin Simitev \url{orcid.org/0000-0002-2207-5789}}
 and Friedrich  H. Busse\aff{2}}

\affiliation{
\aff{1} School of Mathematics and Statistics, University of  Glasgow, Glasgow G12 8SQ, UK
\aff{2} Institute of Physics, University of Bayreuth,  Bayreuth 95440,  Germany}

\begin{document}

\maketitle

\begin{abstract}
Stellar radiative zones are typically assumed to be motionless in standard
models of stellar structure but there is sound theoretical and
observational evidence that this cannot be the case. We investigate
\red{by direct numerical simulations} a three-dimensional and time-dependent model of stellar
radiation zones consisting of an electrically-conductive and
stably-stratified anelastic fluid confined to a rotating spherical
shell and driven by a baroclinic torque. 
As the baroclinic driving is gradually increased a sequence of
transitions from an axisymmetric and equatorially-symmetric
time-independent flow to flows with a strong  poloidal component and
lesser symmetry are found. \red{It is shown that all flow regimes
characterised with significant non-axisymmetric components are capable
of generating} self-sustained magnetic \red{field}. As the value of
the Prandtl number is decreased and the value of the \red{Ekman number
is decreased} flows become strongly time-dependent with
\red{progressively} complex spatial structure and dynamos can be
generated at lower values of the magnetic Prandtl number. 
\end{abstract}

\section{Introduction}

It has long been known theoretically that radiative zones in rotating stars
cannot be in static equilibrium \citep{Zeipel}.
\red{Instead the basic axisymmetric solution
of the problem consists of a differential rotation and a rather weak
meridional circulation driven
by the centrifugal force in the presence of the Coriolis force
\citep{Schwarzschild1947,Busse1982,Spruit1984}}.
In more recent years, fluid motions and the possibility of magnetic
field generation in stably stratified stellar radiative zones have
gained considerable interest owing to increasingly detailed
observations of stellar magnetic fields. To model radiative cores
several numerical studies of baroclinically-driven flows in stably
stratified rotating  spheres have been published, notably
\citep{Rieutord2006a,
  EspinosaLara2013Selfconsistent,Hypolite2014Dynamics,Rieutord2014Dynamics}. 
However, these studies have been restricted to modelling
two-dimensional and steady axisymmetric flows and have been based on
the Boussinesq approximation which does not account for the strong
density variations typical for stably stratified regions in
stars. Also the possibility of magnetic flux generation by dynamo action
has not been considered in these papers. 

These restrictions have been lifted in a recent study by
\cite{SimitevBusse2017} who considered a non-axisymmetric and
time-dependent model of baroclinic flows in rotating spherical fluid
shells based on the anelastic approximation
\citep{Gough1969,BraginskyRoberts1995,LantzFan1999}. \cite{SimitevBusse2017} found a
sequence of bifurcations from simple regular to complex irregular
flows with increasing baroclinicity. Some of the transitions in this
sequence were observed to exhibit hysteretic behaviour. They noted that with increasing
baroclinicity the poloidal component of the flow grows relative to the
dominant toroidal component and thus facilitates magnetic field
generation. \cite{SimitevBusse2017} proceeded to report a
non-decaying dynamo solution and suggested that self-sustained dynamo
action of baroclinically-driven flows may allow for the possibility that
magnetic fields in stably stratified stellar interiors are not
necessarily of fossil origin as is often assumed.

In the present study, we \red{essentially confirm and
significantly extend} the results presented in
\citep{SimitevBusse2017} \red{by summarizing over 90 new hydrodynamic
and hydromagnetic numerical simulations, 62 of which are explicitly
included in the text.} The results of \citep{SimitevBusse2017} were 
restricted to fixed values for all model parameters apart from the
baroclinicity parameter. In particular, the value of the Prandtl
number was fixed to $\Pr=0.1$ which restricts the baroclinicity 
to a fairly modest value and only steady flows were thus reported. In
addition, only a single steady dynamo was presented in 
\citep{SimitevBusse2017} for an unrealistically large value of the
magnetic Prandtl number. Here, we \red{investigate a much larger region
of the parameter space reaching in the directions of} smaller values of the
Prandtl number and of the magnetic Prandtl number and of \red{smaller values
of the Ekman number}. \red{We report one new hydrodynamic instability
and we find essentially time-dependent flows. Another main goal of the
present study is to determine which of the distinct
baroclinically-driven flow states found are capable of generating
self-sustained magnetic field and we are able to establish the
existence of dynamo excitation in all states with non-axisymmetric
flow components. The critical value of the magnetic Prandtl number for
the onset of dynamo action is determined and the dynamo mechanism
is elaborated.}

\section{Mathematical model}

Our model of a stably-stratified stellar radiative zone is based on the anelastic
approximation \citep{Gough1969,BraginskyRoberts1995,LantzFan1999}
which is widely adopted for numerical simulations of convection in
Solar and stellar interiors \citep{JonesBench}.
Accordingly, we consider a perfect gas confined to a spherical shell
rotating with a fixed angular velocity $\varOmega \k$, \red{where $\k$
is the unit vector in the direction of the axis of rotation}.
A positive entropy contrast $\Delta S$ is imposed between its outer
and inner surfaces at radii $r_o$ and $r_i$, respectively.  
{We assume a gravity field proportional to $g/r^2$. To
justify this choice consider the Sun, the star with the best-known
physical properties. The Solar density drops from 150 g/cm$^3$ at the
centre to 20 g/cm$^3$ at the core-radiative zone boundary (at $0.25
R_\odot$) to only 0.2 at the tachocline (at $0.7 R_\odot$). A crude
piecewise linear interpolation shows that most of the mass is
concentrated within the core. In this setting} a hydrostatic
polytropic reference state 
exists with profiles of density, temperature and pressure 
given by the expressions
$$\rhobar = \rho_c\zeta^n,\qquad \Tbar=T_c\zeta, \qquad \Pbar = P_c
\zeta^{n+1},
$$
respectively, where 
\begin{gather*}
\zeta= c_0+\frac{c_1}{r}, \qquad
\zeta_o=\frac{\eta+1}{\eta \exp(\Nrho/n)+1}, \\
c_0=\frac{2\zeta_o-\eta-1}{1-\eta}, \qquad 
c_1=\frac{(1+\eta)(1-\zeta_o)}{1-\eta}^2, 
\end{gather*}
see \citep{JonesBench}.
The parameters $\rho_c$, $P_c$ and $T_c$ are reference values of
density, pressure and temperature at mid-shell. The gas polytropic
index $n$, the density scale height $N_\rho$ and the shell radius
ratio $\eta$ are defined further below. 
\red{Since a rigidly rotating stellar interior can not exist, the
deviation from such a state is described by the baroclinic parameter
$\Z$, introduced below. This parameter is associated with the
centrifugal force which  balances the baroclinic torques \citep{Busse1982}.} 
Following \cite{Rieutord2006a} we neglect the
distortion of the isopycnals caused by the centrifugal force. The
governing anelastic equations of continuity, momentum, energy
(entropy), and magnetic induction assume the form 
\begin{subequations}
\label{govmod2}
\begin{align}
\label{govmod2.01}
{\bm{ \nabla}}{\bm{ \cdot}}\rhobar\u =&\; 0, \\
\label{govmod2.01a}
\quad \quad \bm{\nabla}\bm{\cdot}\B =&\;0, \\
\upartial_t \u + ({\bm{ \nabla}}\times\u)\times\u = &\;-{\bm{ \nabla}}\varPi
-\redmath{\frac{2}{\Ek}}(\k\times\u)+\frac{\R}{\Pr}\frac{{S}}{r^{2}}\ru + \v F_\nu \nonumber \\
\phantom{=}&\;\qquad\quad + \frac{1}{\rhobar} ({\bm
  \nabla}\times\B)\times\B - \Z (\overline{S}+S)\;
\k\times(\r\times\k), 
\label{govmod2.02}\\
\label{govmod2.03}
\upartial_t S + \u{\bm{ \cdot}}{\bm{ \nabla}}{(\overline{S}+S)}
 =&\; \frac{1}{\Pr \rhobar\Tbar} {\bm{ \nabla}}{\bm{ \cdot}}\rhobar\Tbar {\bm{ \nabla}} S
  + \frac{c_1 \Pr}{\R \Tbar}\left(Q_\nu+ \frac{1}{\Pm\rhobar} Q_j\right),\\
\label{govmod2.04}
\upartial_t \B =&\; {\bm \nabla}\times(\u\times\B)+\Pm^{-1} \nabla^2 \B,
\end{align}
\end{subequations}
where $S$ is the deviation from the background entropy profile
$$\overline{S}=\frac{\zeta(r)^{-n}-\zeta_o^{-n}}{\zeta_o^{-n}-\zeta_i^{-n}},$$
$\u$ is the velocity vector, $\B$ is the magnetic flux density, ${\bm{ \nabla}} \varPi$ includes all terms that can
be written as gradients,  and $\r=r \ru$ is the position vector with
respect to the center of the sphere \citep{JonesBench,Simitev2015}.
The viscous force, the viscous heating and the Ohmic heating, given by
$$
\v F_\nu =
(\rho_c/\rhobar){\bm{ \nabla}}{\bm{ \cdot}}\v {\hat S},\qquad 
Q_\nu=\v{\hat S}{\bm{ :}}\v e, \qquad 
Q_j=(\bm{ \nabla}\times\B)^2, 
$$
respectively, are defined in terms of the deviatoric stress tensor   
$$\hat S_{ij}=2\rhobar(e_{ij}-e_{kk}\delta_{ij}/3), \qquad 
e_{ij}=(\upartial_i u_j +\upartial_j u_i)/2,$$
where double-dots (\textbf{:}) denote a Frobenius inner product, and
$\nu$ is a constant viscosity. The governing equations \eqref{govmod2}
have been non-dimensionalized with the shell thickness $d=r_o-r_i$ as unit of
length, $d^2/\nu$ as unit of time, $\nu \sqrt{\mu_0\rho_c}/d$ as a
unit of magnetic
induction, and $\Delta S$, $\rho_c$ and
$T_c$ as units of entropy, density and temperature, respectively. 
The system is then characterized by eight dimensionless parameters:
the radius ratio, the polytropic index of the gas, the density scale
number, the Prandtl number, the magnetic Prandtl number, the Rayleigh
number, the Ekman number and the baroclinicity parameter
\begin{gather*}
\eta=r_i/r_o, \qquad 
n, \qquad 
N_\rho=\ln\big(\rhobar(r_i)/\rhobar(r_o)\big), \qquad
\Pr={\nu}/{\kappa}, \qquad \\
\Pm ={\nu}/{\lambda}, \qquad
\R=-{gd^3\Delta S}/({\nu\kappa c_p}),\qquad
\redmath{\Ek = {\nu}/(\varOmega d^2)},\qquad 
\Z={\varOmega^2 d^4 \Delta S}/({\nu^2 c_p}),
\end{gather*} respectively,
where $\kappa$ is a constant entropy diffusivity, $\lambda$ and
$\mu_0$ are the magnetic diffusivity and permeability, and $c_p$ is the
specific heat at constant pressure. {Note, that the Rayleigh
number assumes negative values in the present problem since the
basic entropy gradient is reversed with respect to the case of
buoyancy driven convection.}

The poloidal-toroidal decomposition
\begin{gather*}
\rhobar \vec u = {\bm{ \nabla}} \times ( {\bm{ \nabla}} \times \ru
r\pol) + {\bm{ \nabla}} \times \ru r^2 \tor,\\
\vec B = \bm{ \nabla} \times  ( \bm{ \nabla} \times \ru \polB) + \bm{ \nabla} \times
\ru \torB,
\end{gather*}
is used to enforce the solenoidality of the mass flux $\rhobar \u$ and
of the magnetic flux density. 
This has the further advantages that the pressure gradient can be  
eliminated and scalar equations for the poloidal and the toroidal
scalar fields, $v$ and $w$, are obtained by taking 
$\ru{\bm{    \cdot}}{\bm{ \nabla}}{\times{\bm{ \nabla}}\times}$
and $\ru{\bm{ \cdot}}{\bm{ \nabla}}\times$ 
of equation \eqref{govmod2.02}. Similarly 
equations for the poloidal and the toroidal
scalar fields, $h$ and $g$, are obtained by taking 
$\ru{\bm{ \cdot}}{\bm{ \nabla}}\times$ and $\ru{\bm{ \cdot}}$ 
of equation \eqref{govmod2.04}. Except for the term
with the baroclinicity parameter $\Z$ the resulting equations are
identical to those described by \cite{Simitev2015}.    
{Assuming that entropy fluctuations are damped by convection in the
region above $r=r_o$ we choose the boundary condition}
\begin{subequations}
\label{BC}
\begin{gather}
  S=0 \quad
\text{at } \quad r
=\left\{\begin{array}{l}
 r_i\equiv\eta/(1-\eta)\,,\\[0.1em]
 r_o\equiv1/(1-\eta)\,,
  \end{array}\right.
\intertext{while the inner and the outer boundaries of the shell are
  assumed
stress-free and impenetrable for the flow}
\pol = 0, \quad
\upartial_r^2 \pol = \frac{\rhobar'}{\rhobar r}\upartial_{r}{}
(r\pol), \quad
\upartial_r (r \tor) = \frac{\rhobar'}{\rhobar} \tor \quad
\text{at } \quad r
=\left\{\begin{array}{l}
 r_i\,,\\[0.1em]
 r_o\,.
  \end{array}\right.
\intertext{The boundary conditions for the magnetic field are derived from the
assumption of an electrically insulating external region. The
poloidal function $\polB$ is then matched to a function $\polB^{(e)}$,
which
describes an external  potential field,}
\torB =0, \quad  \polB-\polB^{(e)} = 0, \quad \partial_r
(\polB-\polB^{(e)})=0 \quad
\text{at } \quad r
=\left\{\begin{array}{l}
 r_i\,,\\[0.1em]
 r_o\,.
  \end{array}\right.
\end{gather}
\end{subequations}

\section{Methods for numerical solution}
A pseudo-spectral method described by \cite{Tilgner1999} was employed
for the \red{direct numerical simulation} of problem (\ref{govmod2}--\ref{BC}). 
We adapted a code developed and used by us for a number of years
\citep{Busse2003a,Simitev2011b,Simitev2015} that has been extensively
benchmarked for accuracy \citep{Marti2014,Simitev2015,Matsui2016}.
Adequate numerical resolution for the simulations was chosen as
described in \citep{Simitev2015}.        
To analyse the properties of the solutions we
decompose the kinetic energy density into poloidal and toroidal
components \red{(respectively denoted by subscripts as in $X_p$ and
$X_t$ in equations \eqref{Engs} below and where $X$ denotes an
appropriate quantity)} and further into mean (axisymmetric) and
fluctuating (nonaxisymmetric) components \red{(respectively denoted by
  bars and tildes as in $\overline X$ and $\widetilde X$ below)}  
and into equatorially-symmetric and equatorially-antisymmetric
components \red{(respectively denoted by superscripts as in $X^s$ and $X^a$ below)}, 
\begin{align}
\overline E_p = \overline E_p^s +\overline E_p^a &=\langle \big({\bm{ \nabla}} \times ( {\bm{ \nabla}} \redmath{(\overline\pol^s+\overline\pol^a)} \times \vec r)
\big)^2/(2\rhobar)  \rangle, \nonumber\\
\label{Engs}
\overline E_t = \overline E_t^s +\overline E_t^a& = \langle \big({\bm{\nabla}} r \redmath{(\overline\tor^s+\overline\tor^a)} \times \vec r \big)^2
/(2\rhobar)  \rangle,  \\
\widetilde E_p = \widetilde E_p^s +\widetilde E_p^a&= \langle
\big({\bm{ \nabla}} \times ( {\bm{ \nabla}} \redmath{(\widetilde
  \pol^s+\widetilde \pol^a)} \times \vec r)  \big)^2/(2\rhobar) \rangle, \nonumber\\
\widetilde E_t = \widetilde E_t^s +\widetilde E_t^a&= \langle \big(
           {\bm{ \nabla}} r \redmath{(\widetilde \tor^s+\widetilde
             \tor^a)} \times\vec r \big)^2 /(2\rhobar) \rangle, \nonumber
\end{align}
where  angular brackets $\langle\,\, \rangle$ denote averages 
over the volume of the spherical shell. \red{Since in our code the spectral 
representation of all fields $X$ is given by the set of coefficients
$\{X_l^m\}$ of their expansions in spherical harmonics $Y_l^m$, it is
easy to extract the relevant components, i.e. coefficients 
with $m=0$ and with $m\ne0$ represent axisymmetric and nonaxisymmetric
components, respectively, while coefficients with even $(l+m)$ and with
odd $(l+m)$ represent equatorially-symmetric and
equatorially-antisymmetric components, respectively.} The magnetic
energy density is similarly decomposed into components.

\section{Parameter values and initial conditions}
In the simulations presented here fixed values are used for some
governing parameters, namely 
\begin{equation}
\label{fixedparms}
\eta=0.3,\qquad n=2, \qquad N_\rho=2, \qquad \R=-5\times 10^4.
\end{equation}
The value for the shell thickness represents a stellar radiative 
core geometrically similar to that of the Sun. 
The values for $n$ and $N_\rho$ are not significantly different from
current estimates for the solar radiative zone, $n=1.5$ and $\Nrho =
4.6$. 
The Rayleigh number is set to a negative value in order to model a
convectively stable configuration.

A large variation is known to exist in the rates of stellar rotation,
e.g.~\citep{Saders2013}. Here, we use three different values of the
\red{Ekman} number, namely \red{$\Ek=2/300$, $1/300$ and $1/500$}, selected so that the
effect of rotation is strong enough to govern the dynamics of the
system, but not too strong to cause a significant increase in
computational expenses; similar values are used e.g.~by
\cite{Simitev2015} and in the cases F1--4 of \citep{Kapyla2017}. This
variation in the values of the \red{Ekman} number allows to observe the
effect of rotation.  

Unresolved subgrid-scales in convective envelopes are typically
modelled by the assumption of approximately equal turbulent eddy
diffusivities. As a consequence the choice of $\Pr=1$ is often made in
the modelling and simulation literature \citep{Miesch2015}. 
However, estimates of the Prandtl number values based on molecular
diffusivities are minute and it is thus unlikely that eddy
diffusivities could increase the effective Prandtl number to a value
of the order of unity. Furthermore, in a stably stratified system turbulence
is expected to be anisotropic \citep{Zahn1992}. With this in mind
we have decreased the value of the Prandtl number
from $0.1$ used in \citep{SimitevBusse2017} to 
smaller values, namely $\Pr=\redmath{0.08}$, $0.05$ and $0.03$. This
variation in the value of $\Pr$ suffices to produce pronounced
differences in the simulation results. Further reductions of $\Pr$
prove to be difficult numerically. 

\red{The molecular values of the magnetic Prandtl number are estimated
to  increase from  about $10^{-5}$ at the surface of the Solar
convection zone to  about $10^{-1}$ at its bottom
\citep{Rieutord2010,Brandenburg2005}. Invoking the eddy
diffusivity argument mentioned above, it may be expected that the
effective values of the magnetic Prandtl number are somewhat further 
increased by turbulent mixing. Considering this, the value of
$\Pm=1.5$ for which a dynamo is demonstrated in the present paper is
certainly large but, perhaps not excessively large.
We remark, that it is possible to decrease $\Pm$ further as
discussed in relation to Figure \ref{fig:060new} later in the
paper. With the aim of establishing the possibility of self-sustained
dynamo action in all of the baroclinically-driven flow states found
(see below) we also report simulations with values of $\Pm$ larger
than 1.5.
Finally, values similar to the ones used by us are also used by other
numerical simulations reported in the literature e.g.~many of the cases in
\citep{Kapyla2017} have values significantly  greater than unity even
though theirs is a model of the Solar convection zone where $\Pm$ is
estimated to be smaller than in the radiative zone.}       

\begin{figure}
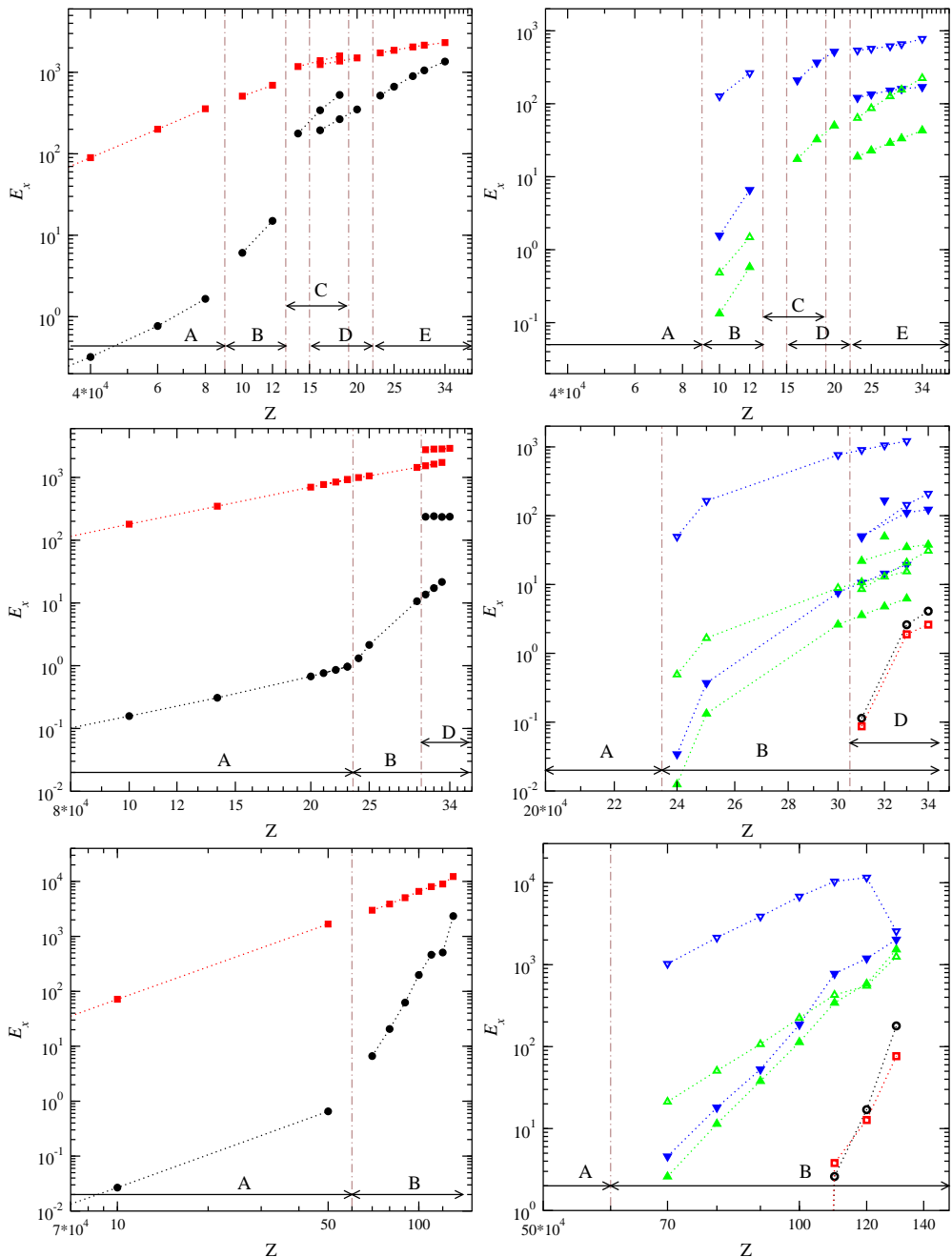

\begin{center}
\mypsfrag{Ex}{$E_x$}
\mypsfrag{z}{\hspace{0mm}$\Z$}
\mypsfrag{ratio}{ratios}
\epsfig{file=Fig04p005t03a.eps,width=0.48\textwidth,clip=}
\epsfig{file=Fig04p005t03b.eps,width=0.483\textwidth,clip=}
\epsfig{file=Fig04p005t06a.eps,width=0.48\textwidth,clip=}
\epsfig{file=Fig04p005t06b.eps,width=0.483\textwidth,clip=}
\epsfig{file=Fig04p003t1a.eps,width=0.48\textwidth,clip=}
\epsfig{file=Fig04p003t1b.eps,width=0.483\textwidth,clip=}
\end{center}
\caption{Time-averaged kinetic energy densities as functions of baroclinicity
$\Z$ for parameter values \eqref{fixedparms} and $\Pr=0.05$,  \red{$\Ek=2/300$} (top
  two panels), $\Pr=0.05$,  \red{$\Ek=1/300$}  (middle two panels), and 
$\Pr=0.03$,  \red{$\Ek=1/500$} (bottom two panels), 
Full and empty symbols indicate equatorially-symmetric and -asymmetric
energy components, respectively. Black  circles, red squares, green
triangles-up and blue  triangles-down indicate the energy components
$\overline E_p^{s,a}$,  $\overline E_t^{s,a}$, $\widetilde E_p^{s,a}$,
$\widetilde E_t^{s,a}$, respectively. Axially-symmetric and
axially-asymmetric components are plotted in the left and the right
panels, respectively. Vertical dash-dotted lines 
indicate transition points. The ranges over which distinct
states are observed are indicated by arrows near the bottom
abscissa, with some states co-existing as indicated. Energy components not shown are at
least 10 orders of magnitude smaller than the ones shown. 
(Colour online)
}
\label{fig:020}
\end{figure}

With these choices of parameter values, we find it convenient to
organize and present our results in terms of several sequences of
cases in which the \red{parameter} $\Z$ is increased in small steps
\red{to study radiative zones of various degrees of baroclinicity}
while all other parameter values in a sequence are kept fixed, see for
instance Figure \ref{fig:020}. We remark that the strength of the
baroclinic forcing, measured by $\Z$, is limited from above so that  
$$
\Z < (1-\eta)^3\; |\R|/\Pr.
$$
This restriction guarantees that the apparent gravity does not point
outward such that the model explicitly excludes the well-studied
case of buoyancy-driven thermal convection.

Initial conditions of no fluid motion are used at vanishingly small
values of $\Z$, while at finite values of $\Z$ the closest
equilibrated neighbouring case is used as initial condition to help
convergence and reduce transients. To ensure that transient effects
are eliminated from the sequences presented below, all solutions have
been continued for at least 15 time units.  Similarly, several dynamo
cases have been started with small random seeds for the magnetic
field, while most other cases were subsequently started from
neighbouring simulations with equilibrated dynamos.

\begin{figure}
\begin{center}
\epsfig{file=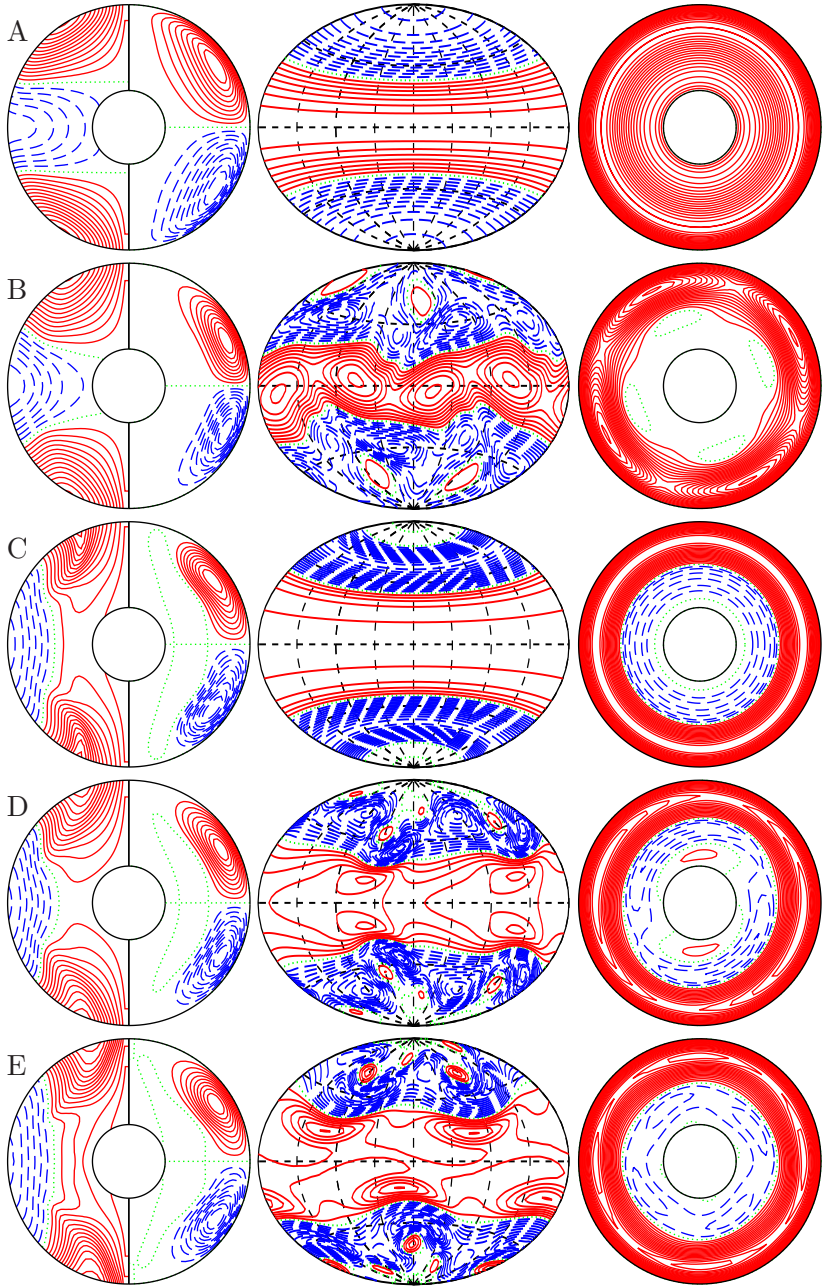,width=0.8\textwidth,clip=}
\end{center}
\caption{
  Flow structures with increasing baroclinicity  $\Z \times 10^{-4} =
  1$, $12$, $18$, $18$ and $30$ from  top to bottom for  $\Pr=0.05$,
  \red{$\Ek=2/300$} and fixed values \eqref{fixedparms}.
  Bistability
  occurs at $\Z \times 10^{-4} = 18$. 
  The first  plot in each row shows isocontours of
  $\overline{u}_\varphi$ (left half) and streamlines
  $r\sin\theta(\upartial_\theta \overline{\pol})=$ const. (right half)
  in the meridional plane. The second plot shows
  isocontours of $u_r$ at $r=r_i+0.7$ \red{maped to the spherical
    surface using isotropic Aitoff projection}. 
The third plot shows
  isocontours of $u_r$ in the equatorial plane. The isocontours are equidistant with positive
  isocontours shown by solid lines,   negative isocontours shown by
  broken lines and the zeroth isocontour shown by a dotted line in each plot.
  \red{All contour plots are snapshots at a fixed representative  moment in time.}  
  \red{Letters at each row denote corresponding flow states as
    indicated in Figure \ref{fig:020}.}
(Colour online)
}
\label{fig:010}
\end{figure}

\begin{figure}
\begin{center}
\epsfig{file=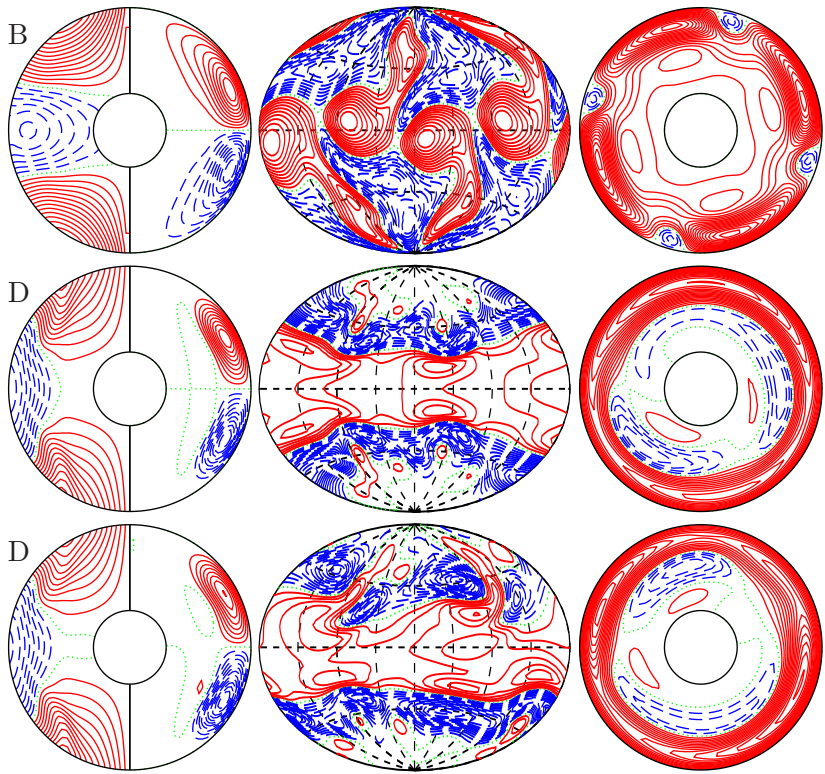,width=0.8\textwidth,clip=}
\end{center}
\caption{
  Flow structures with increasing baroclinicity  $\Z \times 10^{-4} =
  30$, $32$, $34$ for $\Pr=0.05$, \red{$\Ek=1/300$} and fixed values
  \eqref{fixedparms}. 
  \red{The same quantities are plotted in each row as in Figure
  \ref{fig:010} and letters on the left side denote the corresponding flow state as in
  Figure \ref{fig:020}.} 
(Colour online)
}
\label{fig:011}
\end{figure}

\begin{figure}
\begin{center}
\epsfig{file=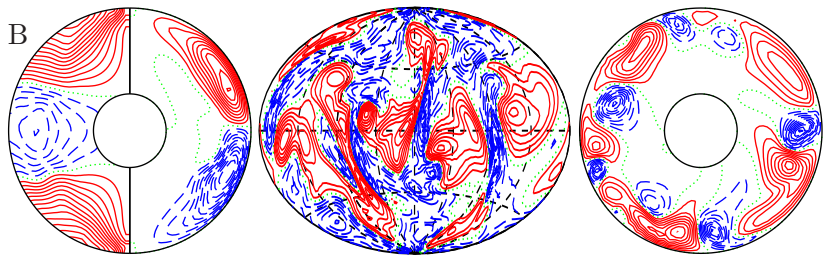,width=0.8\textwidth,clip=}
\end{center}
\caption{
  Flow in the case $\Z \times 10^{-4} = 120$, for $\Pr=0.03$,
  \red{$\Ek=1/500$} and fixed values \eqref{fixedparms}. 
  \red{The same quantities are plotted in each row as in Figure
  \ref{fig:010} and letters denote the corresponding flow state as in
  Figure \ref{fig:020}.} 
(Colour online)
}
\label{fig:012}
\end{figure}

\begin{figure}
\begin{center}
\hspace*{-2.6mm}
\rule[-.3\baselineskip]{0pt}{70mm}
\mypsfrag{A}{$\overline E_p$}
\mypsfrag{B}{$\overline E_t$}
\mypsfrag{C}{$\widetilde E_p$}
\mypsfrag{D}{$\widetilde E_t$}
\epsfig{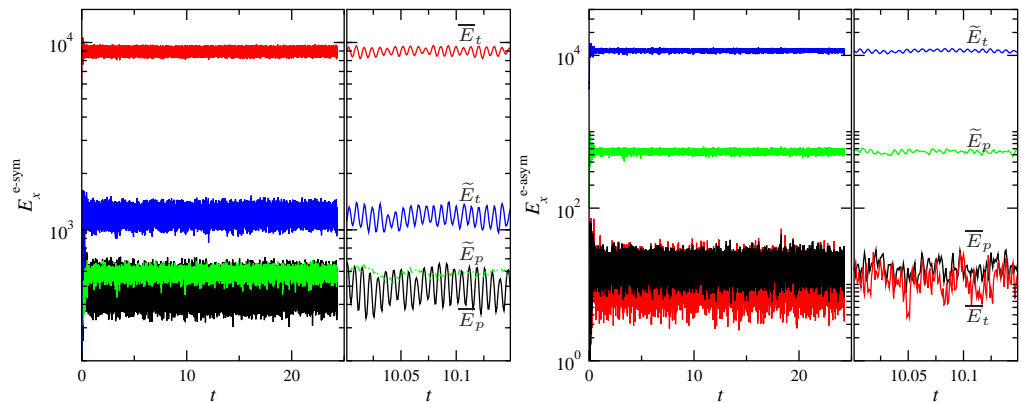}
\end{center}
\caption{
Time series of kinetic energy densities in the case shown in Figure
\ref{fig:012}. Equatorially symmetric components are shown in the left
two panels and equatorially symmetric components are shown in the right
two panels. The second panel in each group shows an enlarged segment
of the time axis.
The  components $\overline E_p$,  $\overline E_t$, $\widetilde
E_p$, $\widetilde E_t$, are labelled in the
plot and are also indicated by black lines, red lines, green lines and blue lines, respectively.} 
\label{fig:012engs}
\end{figure}

\section{Instabilities of baroclinically-driven flows}

The baroclinically-driven problem is invariant under a group of symmetry
operations including rotations about the polar axis
i.e.~invariance with respect to the coordinate transformation
$\varphi\to\varphi+\alpha$, reflections in the equatorial plane
$\theta \to \pi-\theta$, and translations in time $t \to t+a$, see
\citep{Gubbins1993}. 
As baroclinicity $\Z$ is increased the available symmetries of the
solution are broken resulting in a sequence of states ranging from
simpler and more symmetric flow patterns to more complex flow patterns of lesser
symmetry. In this respect the system resembles Rayleigh-B´enard
convection \citep{Busse2003}.  

Three sequences of non-magnetic simulations with gradually increasing
values of the baroclinicity $\Z$ and fixed values of the other 
parameters were obtained and are shown in Figure \ref{fig:020}. 
The first sequence is the most detailed one and allows us to capture the
transitions that occur as the flow is more strongly driven while the
other two sequences allow us to describe the effects of the variation
in \red{$\Ek$} and $\Pr$. 

The sequence with $\Pr=0.05$ and \red{$\Ek=2/300$} is illustrated in the top two
panels of Figure \ref{fig:020} and in Figure \ref{fig:010}.
The sequence starts with the basic axisymmetric, equatorially symmetric and
time-independent state with a dominant wave number $k=1$ in the
radial direction labelled \textbf{A} in Figures \ref{fig:020} and \ref{fig:010}. 
An instability occurs at about {$\Z=9 \times 10^4$} leading to a state
\textbf{B} characterized by a dominant azimuthal wavenumber 
$m=2$ in the expansion of the solution in spherical harmonics $Y_l^m$,
i.e.~{while the axisymmetry is broken,} a symmetry holds with
respect to the transformation
$\varphi\to\varphi+\pi$. {Simultaneously, the symmetry about  
the equatorial plane is also broken in state \textbf{B}.
At $\Z= 13\times 10^4$ a further transition to a pattern
labelled \textbf{C} occurs. This state looks much like state
\textbf{A}, i.e. it is axisymmetric, equatorially symmetric
and time-independent but differs from state \textbf{A} in that it
keeps a dominant radial wave number $k=2$. 
State \textbf{C} ceases to exists above $19 \times 10^4$.
At $\Z= 22 \times 10^4$ a further transition to a pattern labelled
\textbf{E} occurs. 
State \textbf{E} seems to be related to state \textbf{B} in a way similar
as state \textbf{C} is related to state \textbf{B}.
Similarly to \textbf{B}, state \textbf{E} has a $m=2$ azimuthal
symmetry and is asymmetric with respect to the equatorial plane.
\textbf{E} differs from \textbf{B}, however, in that a dominant radial wave number $k=2$ develops.
Remarkably, state \textbf{C} coexists with a  pattern \textbf{D}
that can be found for a range of values of $\Z$ from $15 \times 10^4$ to
$22 \times 10^4$. Like \textbf{C}, state \textbf{D} is equatorially
symmetric, has a dominant radial wave number $k=2$  and is
time-independent. Unlike \textbf{C} it is not axisymmetric, but exhibits a
$m=2$ structure.
Which one of the coexisting patterns \textbf{C} and \textbf{D} will be
found in a given numerical simulation depends on the 
initial conditions.

The \red{Ekman number is decreased} to \red{$\Ek=1/300$} in our second
sequence while we keep $\Pr=0.05$. The sequence is illustrated in the middle two panels
of Figure \ref{fig:020} and in Figure \ref{fig:011}. States \textbf{C}
and \textbf{E} are not observed. State \textbf{B} becomes rather more
pronounced with patches detaching from each other and with arms
shooting towards the poles. State \textbf{D} loses its two-fold
azimuthal symmetry $m=2$. In addition, state \textbf{D} becomes
time-periodic at the largest values of $\Z$, e.g. at $\Z=34\times10^4$
shown in the last row of Figure \ref{fig:011}.

The Prandtl number is decreased to $\Pr=0.03$ and the \red{Ekman
  number is further decreased} to \red{$\Ek=1/500$} in our third sequence. The
kinetic energy components are plotted in the bottom two panels of
Figure \ref{fig:020}. In comparison with the preceding sequence, state
\textbf{D} is not observed. For values of the baroclinicity larger
than $\Z=110\times10^4$ the spatial structure of state \textbf{B}
becomes irregular as shown in Figure \ref{fig:012} and exhibits a chaotic
time-dependence as illustrated by the time series of its kinetic
energy components in Figure \ref{fig:012engs}.

\red{A summary of the symmetry properties of all distinct flow states
identified is listed in Table \ref{Table02} as is their capability
of self-sustained dynamo action, see next section.} 

\red{We wish to close this section by a brief comparison of
baroclinically driven flows with the extensively studied problem of
convective flows in rotating systems. Typical convection driven
finite-amplitude instabilities are described for instance by
\citep{SunSchubert1993,Simitev2003,BusseSimitev2005} and it is
apparent that the baroclinically driven flows described in the article
are essentially different. One feature that appears similar at first sight
is the finding that baroclinically driven flows similarly to
buoyancy-dominated convective turbulent flows (or equivalently slowly
rotating convection) are characterised by retrograde (anti-solar)
differential rotation. However, in rotating turbulent convection
anti-solar rotation occurs after a sharp transition from prograde
(solar) rotation as the buoyancy is gradually increased
\citep{Simitev2015}, see also \citep{Gastine2013,Kpyl2014}. Moreover, convection in the
solar rotation regime is characterised by elongated convective
columns, known as ``Busse columns'', while convection in the anti-solar
regime is strongly disorganised, see \citep{Simitev2015}. Furthermore,
anti-solar rotation is reversed to solar rotation in the presence of
magnetic field \citep{Simitev2015} and other effects of the magnetic
field on this transition are reported in \citep{Fan2014,Karak2015}. In contrast, in baroclinically
driven flows anti-solar rotation occurs at all values of the
baroclinicity starting from the basic flow state \textbf{A}, the
baroclinic flow is regular and well-organised, and the magnetic fields
generated by baroclinic dynamos do not reverse anti-solar rotation as
seen in the figures of section \ref{dynamos} below. }

 \begin{table*}
 \begin{center}
 \begin{tabularx}{0.7\columnwidth}{l@{\hspace{10mm}}l@{\hspace{5mm}}l@{\hspace{5mm}}l@{\hspace{5mm}}l}
 \toprule
         & Equatorial&  Azimuthal & Radial  & \red{Dynamo} \\
         & symmetry & symmetry & structure  & \red{capability}\\
 \midrule
 State \textbf{A}  & yes & yes    & 1-roll & \red{no} \\
 State \textbf{B}  & no  & 2-fold & 1-roll & \red{yes} \\
 State \textbf{C}  & yes & yes    & 2-roll & \red{no} \\
 State \textbf{D}  & yes & 2-fold & 2-roll & \red{yes} \\
 State \textbf{E}  & no  & 2-fold & 2-roll & \red{yes} \\
 \toprule
 \end{tabularx}
 \caption{Summary of \red{symmetry} properties and \red{dynamo
     capability} of states in the case $\Pr=0.05$, \red{$\Ek=2/300$} and fixed values \eqref{fixedparms}.} 
 \label{Table02}
 \end{center}
 \end{table*}

\section{Dynamos generated by baroclinically-driven flows}
\label{dynamos}

The possibility of dynamos generated in stably stratified stellar
radiative regions has received only limited support in the literature,
e.g.~\citep{Braithwaite2006}, because it is well known that dynamo
action does not
exist in a spherical system in the absence of a radial component of
motion \citep{BullardGellman2954,KaiserBusse2017}. The latter is, indeed, rather weak
in the basic state \textbf{A} of low poloidal kinetic energy
{as} discussed above. However, with increasing baroclinicity
$\Z$, the  growing radial component of the velocity field strengthens
the likelihood of dynamo action.
\red{
Indeed, the existence of dipolar dynamos sustained by the
equatorially-symmetric flow state \textbf{D} was demonstrated in
\citep{SimitevBusse2017} for $\Pr=0.1$, \red{$\Ek=2/300$} and $\Pm=16$. 
Below we investigate further the existence of dynamo action in 
all other baroclinically-driven flow states described in the
preceding section.}   

A quadrupolar dynamo solution sustained by a baroclinically-driven flow in state
\textbf{B} at  $\Pr=0.03$, \red{$\Ek=1/500$} and $\Z=100 \times 10^4$ and
for a magnetic Prandtl number value $\Pm=4$ is presented in Figure
\ref{fig:041}.
This is a steady dynamo. 
The ratio of poloidal to toroidal magnetic energy is
$E_p^\text{magn}/E_t^\text{magn}=0.033$, the ratio of $E^\text{magn}_\text{dipole}/E^\text{magn}_\text{quadrupole}=0.596$ and the ratio of magnetic to
kinetic total energy is $E^\text{magn}/E^\text{kin}=0.009$. 
The magnetic field has a weaker dipolar component and a large-scale
quadrupolar topology that does not change in time.
The surface structure of the magnetic field is characterised by
relatively strong
inward magnetic flux at both poles. Four localised patches of inward
magnetic flux appear in the equatorial region. 
The azimuthally-averaged toroidal field shows two large scale strong
flux tubes near the pole. The azimuthally-averaged poloidal field is
rather remarkable in that it remains almost completely confined to the
spherical shell giving rise to an ``invisible dynamo''. 

This invisibility, however, does not persist as illustrated by a more
strongly-driven quadrupolar dynamo at $\Z=110 \times 10^4$ and $\Pm=2$
shown in Figure \ref{fig:042}. The dynamo exhibits regular
oscillations in which the quadrupole component reverses polarity. 

\begin{figure*}
\begin{center}
\begin{tabular}{cc}
\raisebox{6cm}{\textbf{B} }&
\epsfig{file=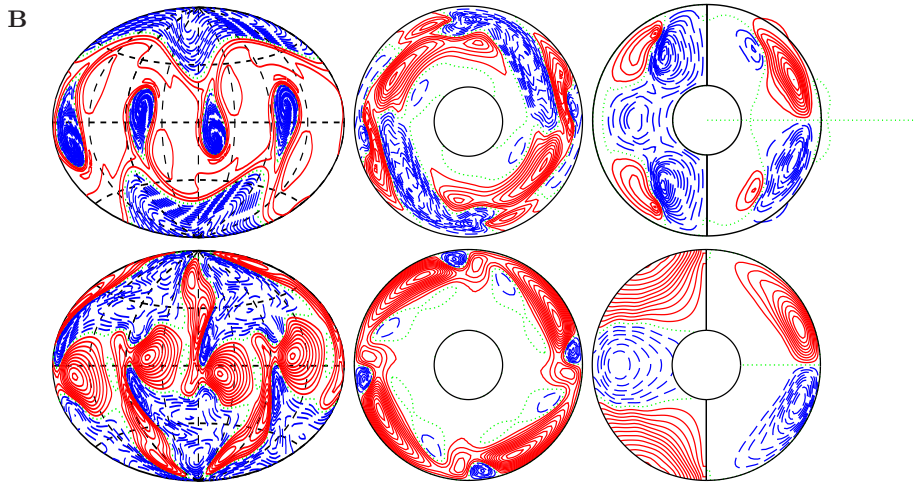,width=0.85\textwidth,clip=}\\
\end{tabular}
\end{center}
\caption{A dynamo solution \red{for a flow in state \textbf{B}} in the
case $\Z \times 10^{-4} = 100$, with $\Pr=0.03$, \red{$\Ek=1/500$}, $\Pm=4$
and values \eqref{fixedparms}. 
First row: 
The first plot shows isocontours of radial magnetic field ${B}_r$ at 
$r=r_o+0.1$ \red{in isotropic Aitoff projection.}
The second plot shows contours of $B_r$ in the equatorial plane.
The third plot shows isocontours of $\overline{B}_\varphi$
(left half) and meridional field lines $r \sin\theta\,
\upartial_\theta                                                                      
\overline{h} = $ const. (right half). 
Second row:
The first plot shows isocontours of radial velocity $u_r$
at $r=r_i+0.7$ \red{in isotropic Aitoff projection.}
The second plot shows contours of $u_r$ in the equatorial plane.
The third plot shows  isocontours of
$\overline{u}_\varphi$ (left  half) and streamlines
$r\sin\theta(\upartial_\theta \overline{\pol})=$ const. (right half)
in the meridional plane.
  \red{All contour plots are snapshots at a fixed representative  moment in time.}  
\red{The letter B denotes the corresponding flow state as indicated in Figure \ref{fig:020}.} 
(Colour online)}
\label{fig:041}
\end{figure*}
\begin{figure*}
\begin{center}
\begin{tabular}{cc}
\raisebox{5.9cm}{\textbf{B} }&
\epsfig{file=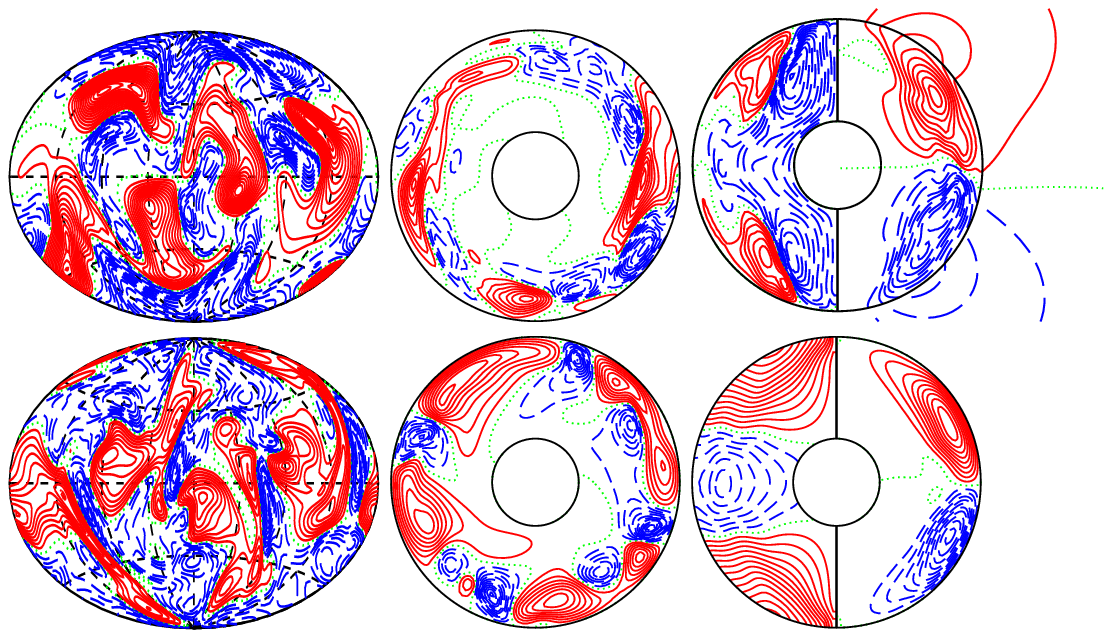,width=0.84\textwidth,clip=}\\
\end{tabular}
\end{center}
\caption{A dynamo solution \red{for a flow in state \textbf{B}} in the case $\Z \times 10^{-4} = 110$,
  $\Pr=0.03$, \red{$\Ek=1/500$}, $\Pm=2$ and values \eqref{fixedparms}.
The same quantities are plotted as in Figure \ref{fig:041} 
\red{and the letter B denotes the corresponding flow state as indicated in  Figure \ref{fig:020}.} 
(Colour online)}
\label{fig:042}
\end{figure*}

\begin{figure*}
\begin{center}
\begin{tabular}{cc}
\raisebox{5.7cm}{\textbf{D} }&
\epsfig{file=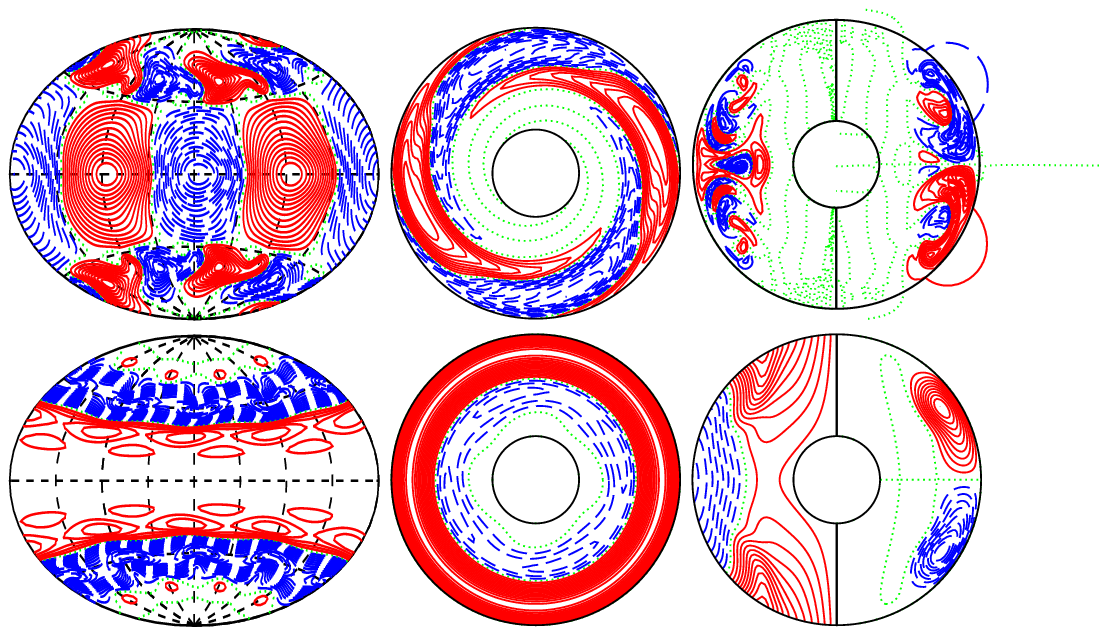,width=0.81\textwidth,clip=}\\
\end{tabular}
\end{center}
\caption{\red{A dynamo solution \red{for a flow in state \textbf{D}} in the case $\Z \times 10^{-4} = 32$, $\Pr=0.05$, \red{$\Ek=1/300$}, and
  $\Pm=5$  and values \eqref{fixedparms}.
The same quantities are plotted as in Figure \ref{fig:041}
\red{and the letter D denotes the corresponding flow state as indicated in  Figure \ref{fig:020}.} 
(Colour online)}}
\label{fig:044}
\end{figure*}
\begin{figure*}
\begin{center}
\begin{tabular}{cc}
\raisebox{5.8cm}{\textbf{E} }&
\epsfig{file=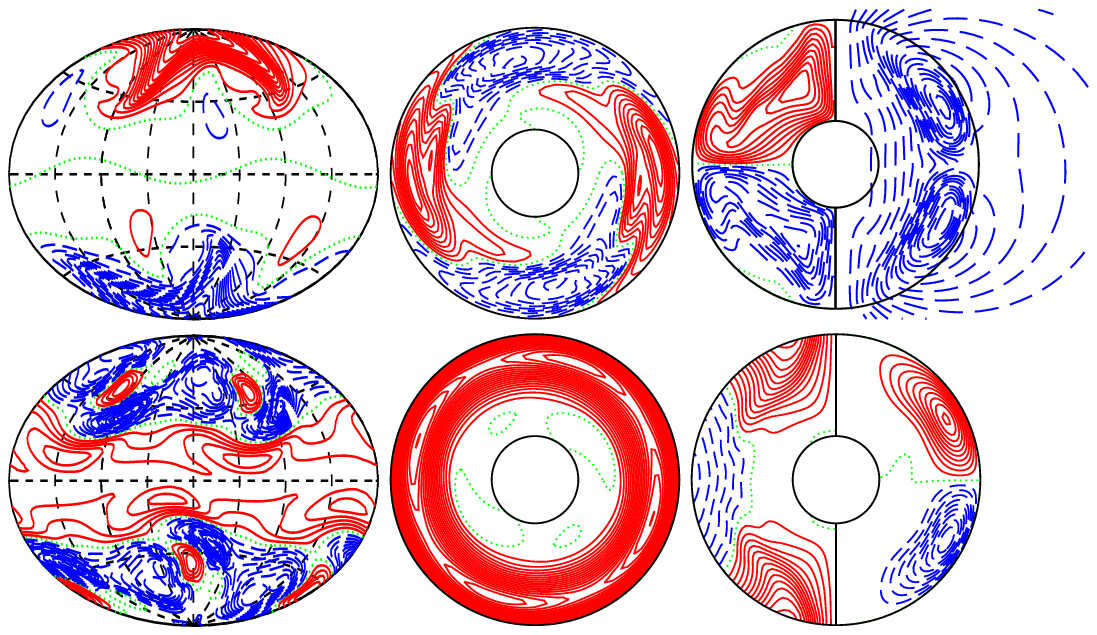,width=0.81\textwidth,clip=}\\
\end{tabular}
\end{center}
\caption{\red{A dynamo solution \red{for a flow in state \textbf{E}} in the case $\Z \times 10^{-4} = 34$, $\Pr=0.05$, \red{$\Ek=2/300$}, and  $\Pm=1.5$  and values \eqref{fixedparms}.
The same quantities are plotted as in Figure \ref{fig:041}
\red{and the letter E denotes the corresponding flow state as indicated in  Figure \ref{fig:020}.} 
(Colour online)}}
\label{fig:043}
\end{figure*}

\red{
A multipolar dynamo solution sustained by a baroclinically-driven flow in state
\textbf{D} at  $\Pr=0.05$, \red{$\Ek=1/300$} and $\Z=32 \times 10^4$ and
$\Pm=5$ is presented in Figure \ref{fig:044}.
This dynamo exhibits a steady time dependence. 
The ratio of poloidal to toroidal magnetic energy is
$E_p^\text{magn}/E_t^\text{magn}=0.384$, the ratio of
$E^\text{magn}_\text{dipole}/E^\text{magn}_\text{quadrupole}=0.0004$
and the ratio of magnetic to kinetic total energy is
$E^\text{magn}/E^\text{kin}=0.003$.  
The spatial structure of the magnetic field of this dynamo is highly
unusual in that it is not dominated by a large scale dipolar component
($Y_1^0$) or quadrupolar component ($Y_2^0$) as most other large scale
dynamos reported to date but rather resembles approximately the
structure of the $Y_6^4$ spherical harmonic function as seen in the
leftmost plot in the upper row of Figure \ref{fig:044}.
Because of the weak $m=0$ contribution to the magnetic field the
azimuthally-averaged components plotted in the rightmost plot in the
upper row of Figure \ref{fig:044} are insignificant.
It is remarkable that in this case the magnetic field
significantly alters the fluid flow pattern as evident by comparison
with the nonmagnetic simulation at identical parameter values 
shown in the second row of Figure \ref{fig:011}, that was 
used as an initial condition for the dynamo run.
The new magnetic flow pattern retains profiles of the differential
rotation and the meridional circulation similar to those of the
nonmagnetic case. However, the dominant azimuthal wave number
increases from 2 to 4 with a stronger contribution of $m=0$. 
None of the other dynamos we report alter their generating flow states
so significantly.
}

A dipolar dynamo generated by a baroclinically-driven flow in state
\textbf{E} for $\Pr=0.05$, \red{$\Ek=2/300$}, $\Z \times = 34\times 10^4$,
and \red{$\Pm=1.5$} is shown in Figure \ref{fig:043}. 
The case has been started from an equilibrated neighbouring case to
help convergence and reduce transients.   
The dynamo solution is steady.
The ratio of poloidal to toroidal magnetic energy is
 $E_p^\text{magn}/E_t^\text{magn}=\red{0.078}$, the ratio of
$E^\text{magn}_\text{dipole}/E^\text{magn}_\text{quadrupole}= \red{5.97}$ and
the ratio of magnetic to 
 kinetic total energy is $E^\text{magn}/E^\text{kin}=\red{0.344}$. 
 Furthermore, the energy density of the magnetic field  $E^\text{magn}$
 is comparable to the kinetic energy density of the poloidal component
 of the velocity field, $E_p^\text{kin}$, with a ratio $E^\text{magn}/E_p^\text{kin}=\red{1.42}$.
 The magnetic field has a weaker quadrupolar component and a
 large-scale dipolar topology with prominent polar magnetic fluxes
 that does not change in time, as shown in  Figure \ref{fig:043},
 resembling in this respect the surface fields of  Ap-Bp stars \citep{Donati2009}. 
 The azimuthally-averaged toroidal field shows a pair of hook-shaped
 toroidal flux tubes largely filling each hemisphere of the  shell. 
 A large scale dipolar component emerges outside of the spherical
 shell. 
 The structure of the fluid flow remains largely unaffected by the
 presence of the magnetic field.

\begin{figure}
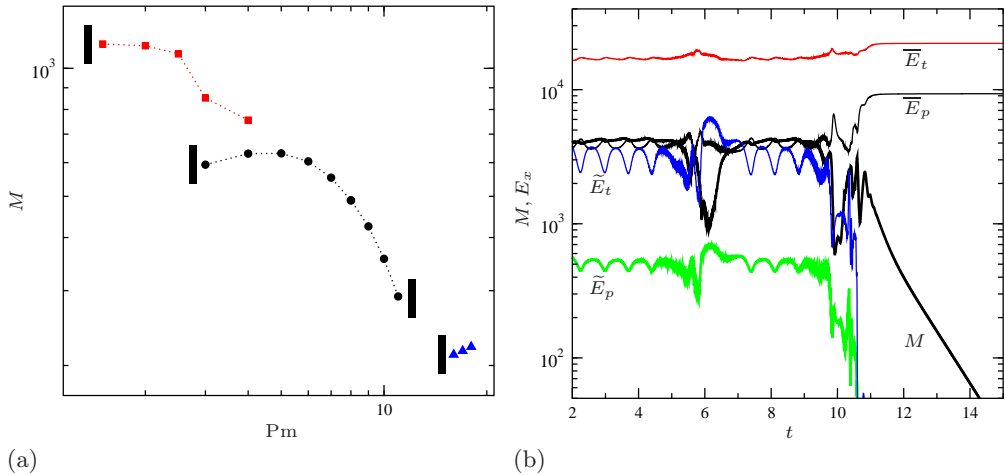

\begin{center}
\mypsfrag{Ex}{$E_x$}
\mypsfrag{Mt}{$M$}
\mypsfrag{Mt,Ex}{$M, E_x$}
\mypsfrag{Pm}{$\Pm$}
\mypsfrag{t}{\hspace{0mm}$t$}
\mypsfrag{Emp}{\hspace{0mm}$\overline E_p$}
\mypsfrag{Emt}{\hspace{0mm}$\overline E_t$}
\mypsfrag{Efp}{\hspace{0mm}$\widetilde E_p$}
\mypsfrag{Eft}{\hspace{0mm}$\widetilde E_t$}
\begin{tabular}{ll}
\epsfig{file=Dynamos.eps,width=0.48\textwidth,clip=}&
\epsfig{file=ae03p008t03r-50000m2p12n2N2z20e4s.eps,width=0.483\textwidth,clip=}\\
\red{(a)} & \red{(b)} \\
\end{tabular}
\end{center}
\caption{\red{
(Colour online)
Left panel (a): Time-averaged total magnetic energy density as a function
of the magnetic Prandtl number for \red{$\Ek=2/300$} in the cases $\Pr=0.1$,
$\Z=17\times10^4$ (blue triangles), $Pr=0.08$, $\Z=20\times10^4$ (black
circles), $P=0.05$, $\Z=32\times10^4$ (red squares). Small black
rectangles indicate the boundary of a region of no dynamo excitation
for the respective sequence. Right panel (b): Components of the
magnetic and kinetic energy densities (as labelled) as a function of
time in the case $Pr=0.08$, $\Z=20\times10^4$, 
\red{$\Ek=2/300$}, $\Pm=12$.  
}}
\label{fig:060new}
\end{figure}

\red{An important issue for application of these results to
astrophysical objects is to determine whether dynamo action persists
at sufficiently small values of the magnetic Prandtl number.
Figure \ref{fig:060new}(a) shows the dependence of the time-averaged
total magnetic energy density on the magnetic Prandtl number for three
sequences of self-sustained dynamo cases with \red{$\Ek=2/300$}. 
In each sequence no active dynamo is found for smaller values of
$\Pm$; thus the smallest value of $\Pm$ is an estimate of the critical
value of $\Pm$ for dynamo onset at fixed values of the other
parameters. In particular, the figure shows that $\Pm_\text{crit}$
decreases as the value of the ordinary Prandtl number $\Pr$ is decreased and the
value of the baroclinicity parameter $\Z$ is increased reducing
$\Pm_\text{crit}$ to $1.5$ at $P=0.05$ and $\Z=32\times10^4$. While it
is numerically challenging to proceed in the direction of stronger
baroclinic driving and smaller ordinary Prandtl number this trend,
which we believe is likely to continue, indicates that baroclinically
driven dynamos at more realistic values of $\Pm$ eventually exist. In this respect baroclinically
driven dynamos seem to behave similarly to both Taylor-Couette dynamos
\citep{Willis2002} and to randomly-forced small-scale dynamos
\citep{Schekochihin2004,Schekochihin2005} where evidence is found that
for fixed other parameter values a critical value of $Pm$ exists below
which dynamo action is not possible. 
Surprisingly, Figure \ref{fig:060new}(a) also shows that there is an
optimal value of $\Pm$ for magnetic field generation,
e.g.~$\Pm_\text{opt}=5$ at $\Pr=0.08$, $\Z=20\times10^4$,  and that
dynamo action is lost when sufficiently large values of $\Pm$ are
used. This appears to be due to a decline of the non-axisymmetric
poloidal and toroidal kinetic energy components, $\widetilde E_p$ and
$\widetilde E_t$, which are most strongly depleted by the magnetic
field. This is related to the specific mechanism of dynamo excitation as
discussed below. }

\red{A common feature of all dynamo-capable baroclinic flow states is
that they exhibit significant non-axisymmetric flow components, $\widetilde E_p$ and
$\widetilde E_t$ as seen in the left panels of Figure \ref{fig:020}. 
These non-axisymmetric flow components appear to be essential for
dynamo action as strikingly illustrated in Figure \ref{fig:060new}(b).
Here a solution for $\Pr=0.08$, $\Z=20\times10^4$, \red{$\Ek=2/300$} and for
$\Pm=12$ is shown. In the initial part of the simulation an
oscillatory dipolar dynamo field is sustained by a flow in state
\textbf{D} which is characterized by significant non-axisymmetric
flow components. Flow state \textbf{D}, however, co-exists with flow
state \textbf{C} which has negligibly small non-axisymmetric flow
components and does not seem to support dynamo action. As  $\widetilde
E_p$ and $\widetilde E_t$ are weakened by the magnetic field a flow
transition from state \textbf{D} to state \textbf{C} occurs after
which the dynamo rapidly decays. 
On their own, the mean components of the flow
which in the basic state \textbf{A} consist of radially-decreasing
differential rotation (``subrotation'') coupled with counterclockwise
meridional  circulation in the northern hemisphere 
do not lead to dynamo excitation as also shown in kinematic dynamo
models with prescribed laminar flows by \citet{Dudley1989}, see also
\citep{Latter2010}. However, since the differential rotation remains the
dominant flow component in all states the baroclinically driven
dynamos reported here are of $\alpha\Omega$ type.  
}

\section{Conclusion}
Evidence of dynamical processes such as differential rotation,
meridional circulation, turbulence, and internal waves in radiative
zones is emerging from helio- and asteroseismology
\citep{TurckChieze2008,Thompson2003,Gizon2010,Chaplin2013,Aerts2010}.
These transport processes are important in the mixing of angular
momentum as well as in the variation of chemical abundances
\citep{Miesch2009,Mathis2013} and are thus critical for the
formulation of a robust stellar evolution theory.   
Further, about 10\% of observed main sequence and pre-main-sequence
radiative stars exhibit detectable surface magnetic fields, 
\citep{Donati2009} posing the question of their origin. 
Thus, fluid motions in radiative zones and their dynamo capability
are of significant interest.

Following our earlier work \citep{SimitevBusse2017} we report
additional direct numerical simulations of non-axisymmetric and
time-dependent flows of anelastic fluids driven by baroclinic torques 
in stably stratified rotating spherical shells -- a system serving as
a simple model of a stellar radiative zone. 
We confirm  that the
general picture described in the latter study persists when values of
the ordinary and the magnetic Prandtl number and of the \red{Ekman number}
are changed up to a factor of 10. We find an additional baroclinic
flow state \textbf{E} and we observe time-dependent behaviour in
states \textbf{B} and \textbf{D}.  \red{The baroclinically driven flows
appear very different from finite-amplitude convective flows in
rotating systems as is apparent from comparisons with published results,
e.g. \citep{SunSchubert1993,Grote2002,Simitev2003}. In particular, the
retrograde differential rotation observed in the present baroclinic
simulations does not seem to be related to the anti-solar rotation known to
characterize turbulent convection in spherical shells rotating at slow  rates \citep{Simitev2015}.}

The increasing spatial and temporal complexity of \red{baroclinic} flow
gives rise to \red{the possibility of dynamo excitation. We have
established that all baroclinically-driven flow states characterised
by significant non-axisymmetric flow components, i.e. states
labelled \textbf{B}, \textbf{D} and \textbf{E}, are capable of generating
self-sustained magnetic fields for an appropriate range of values of
the magnetic Prandtl number.} Dynamo solutions with dipolar,
quadrupolar \red{and multipolar} symmetries are found.
These dynamos provide \red{examples very different from the more
familiar convection-driven dynamos
\citep{Busse2005,BusseSimitev2005,Simitev2012} and are certainly of
general theoretical interest, e.g.} some of them are sustained by
essentially equatorially asymmetric flow fields. \red{We have determined
that a critical value of the magnetic Prandtl number for the onset of 
dynamo excitation $\Pm_\text{crit}$ exists similarly to Taylor-Couette dynamos
\citep{Willis2002}, randomly-forced small-scale turbulent dynamos
\citep{Schekochihin2004,Schekochihin2005} and convectively-driven
dynamos \citep{BusseSimitev2005} and within the range of our numerical
simulations we find that $\Pm_\text{crit}$ decreases with the decrease of the
ordinary Prandtl number and the increase of the baroclinicity
parameter. A surprising new finding is that there exist an ``upper''
critical magnetic Prandtl number such that a shutdown of dynamo excitation
and the decay of the magnetic field occurs at values in excess of
it. This is due to a transition from a flow regime with significant
non-axisymmetric components to a flow regime which is essentially
axisymmetric caused by the presence of the magnetic field itself. 
}

\red{
It must be admitted that our results are not likely to describe even
semi-quantitatively the situation in any star. The choice of
numerically accessible systems is too restricted for detailed
comparisons with astrophysical observations. But we hope that various
features described in this paper can eventually be related to observed
magnetic properties of stars.} 

A baroclinic basic state is known to enhance the tidal instability in
stellar equatorial planes \citep{Kerswell1993,Lebars2006,Vidal2018}. Thus
baroclinically-driven flows could be even more capable of dynamo
action and deserve further study. In terms of future work, it will
be of interest to 
extend the present results to elliptic containers where tidal effects
can be included.  A further line of study is to investigate to what
extent magnetic fields generated by baroclinic driving may be amplified
by a tachocline shear layer at the top of the shell.

\section*{Acknowledgements}
This work was supported by the Leverhulme Trust [grant number RPG-2012-600].

\end{document}